# INTEGRABLE MODELS ON HYPER-ELLIPTIC SURFACES[1]

S. A. Apikyan[2]
Theoretical Physics Department
Yerevan Physics Institute
Alikhanyan Br.st.2, Yerevan, 375036 Armenia

C. J. Efthimiou[3]
Newman Laboratory of Nuclear Studies
Cornell University
Ithaca, NY 14853-5001, USA

**Abstract**

We present an elementary introduction to the construction of integrable models on hyper-elliptic surfaces for non specialists; also, we present some of the details of [1] for the more interested readers.

## 1 Introduction

During the last two decades an essential progress has been achieved in the investigation of integrable quantum field theories. Many models of 2-dimensional Quantum Field Theory (QFT) which admit exact solutions have been constructed on the plane and studied thouroughly:
- Sine-Gordon Model
$$\mathcal{L} = \frac{1}{2}(\partial_\mu \phi)^2 - \frac{m^2}{\lambda}\left(1 - \cos\frac{\sqrt{\lambda}\phi}{m}\right)$$
- Massive Thirring Model
$$\mathcal{L} = \overline{\psi}(i\partial\!\!\!/ - m)\psi - g(\overline{\psi}\gamma^\mu\psi)^2$$
- Bukhvostov-Lipatov Model
$$\mathcal{L} = \overline{\psi}_1(i\partial\!\!\!/ - m)\psi_1 + \overline{\psi}_2(i\partial\!\!\!/ - m)\psi_2 - g(\overline{\psi}_1\gamma^\mu\psi_1)(\overline{\psi}_2\gamma_\mu\psi_2)$$
- ........

In fact, the above list is infinite and growing!

Despite this progress though, our understanding of integrable models on arbitrary Riemann surfaces is still very poor. Such an understanding may be proved essential towards the solution of the challenging problem of quantization of gravity.

In this paper, we present an elementary introduction (and thus non-rigorous) to the notion of hyper-elliptic surfaces, and to the construction of integrable models on these surfaces. Some of the calculations of [1] are also given for the interested reader.

---

[1]Based on a talk given at the MRST 95 meeting by C. E.
[2]e-mail address: apikyan@vx2.yerphi.am
[3]e-mail address: costas@hepth.cornell.edu

# 2 The Notion of Hyper-Elliptic Surfaces

The object under investigation will be the function $w = w(z)$ which satisfies the quadratic equation

$$a_0(z) w^2 + a_1(z) w + a_2(z) = 0 ,\qquad(2.1)$$

where $a_0(z), a_1(z), a_2(z)$ are polynomials in $z$ and $a_0(z) \neq 0$. Using the change of variables

$$\zeta = 2 a_0(z) w + a_1(z) ,\qquad(2.2)$$

we obtain

$$\zeta^2 = P_h(z) ,\qquad(2.3)$$

where $P_h(z) \equiv a_1^2(z) - 4 a_0(z) a_2(z)$ is a polynomial in $z$ of degree $h$. The set $\Gamma$ of pair of points $(z, \zeta)$ such that (2.3) is true is what we call a hyper-elliptic surface. Notice that such a surface has always two branches labelled by the numbers $l = 0, 1$:

$$\zeta^{(l)}(z) = e^{i\pi l} \sqrt{P_h(z)} = e^{i\pi l} \prod_{i=1}^{h} (z - r_i)^{1/2} .\qquad(2.4)$$

For simplicity we shall assume that all $r_i$ are pairwise different.

Let us try to visualize as to what the hyper-elliptic surfaces look like. We start with the simplest possible polynomial:

$$\zeta^2 = z .\qquad(2.5)$$

As every student of complex calculus knows, in this case, $\Gamma$ is represented by two sheets joined at a branch cut running from $0$ to $\infty$.

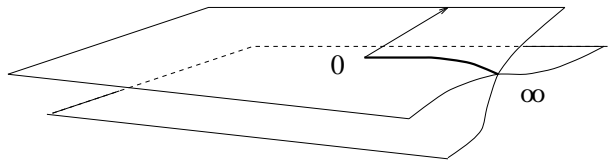

Let us recall now that if we use the stereographic projection, the plane can be seen as a sphere, exactly as seen in the following picture.

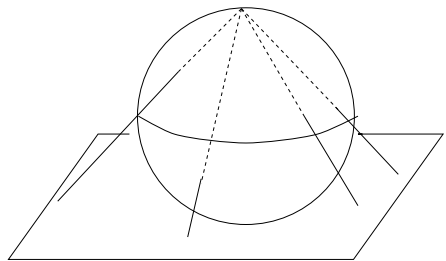

Now, using this information, the surface Γ of the picture 2 can be thought as two spheres each of which carries a branch cut. We think of the two spheres as being made of an elastic material that can be deformed freely in any shape. This is the informal way to speak of a *topological mapping*. Then, we cut the two spheres open along the branch cut creating a hole, deform the edge of each hole to create a small tube and finally join the tubes of the two spheres in the same way we would join the initial Riemann sheets. This sequence of transformations is seen in the picture on the right. Notice that the superposition of the initial two spheres gave a surface which is equivalent to a sphere! In other words, we have created a mapping $S^2 \xrightarrow{w=z^2} S^2$ from a sphere parametrized by $z$ to another sphere $S^2$ parametrized by $w$ which has two branches. This mapping is usually called a *two-sheeted covering of the sphere*.

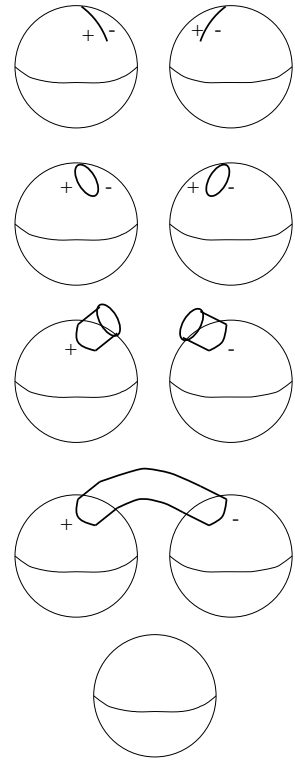

If we now consider the polynomial $P_2 = \alpha\,(z - r_1)(z - r_2)$, an investigation similar to the above shows that nothing new occurs in this case. In particular, the points $r_1, r_2$ are branch points but the point at infinity does not affect the number of a branch. The topological surface we find is again equivalent to a sphere.

Increasing the degree $h$ of the polynomial though, we do discover new surfaces. Lets's take the next simplest case

$$P_3 = \alpha\,(z - r_1)(z - r_2)(z - r_3)\ . \tag{2.6}$$

In this case, we have two Riemann sheets, corresponding to the two solutions of (2.4), joined with two branch cuts running from $r_1$ to $\infty$ and $r_2$ to $r_3$.

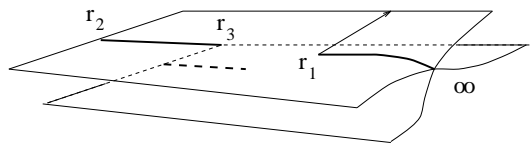

Starting with two spheres and two cuts on each of them now, again cutting open the spheres along the cuts, deforming the edges of the created holes to form tubes and finally connecting the corresponding tubes, we see that the surface $\Gamma$ is now equivalent to the sphere with one handle, i.e. a torus. It is almost obvious now, that increasing the order of the polynomial $P_h(z)$, i.e. increasing the number of roots, we increase the number of cuts on the two Riemann sheets and thus the number of tubes on them. Since every time time we glue two tubes belonging on different spheres we create a new handle, we conclude that we can create Riemann surfaces with any number of handles $g$ (called the *genus* of the surface).

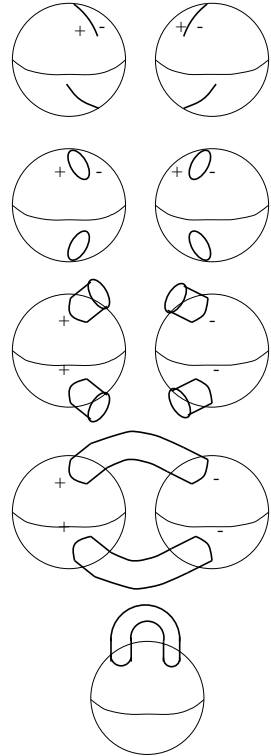

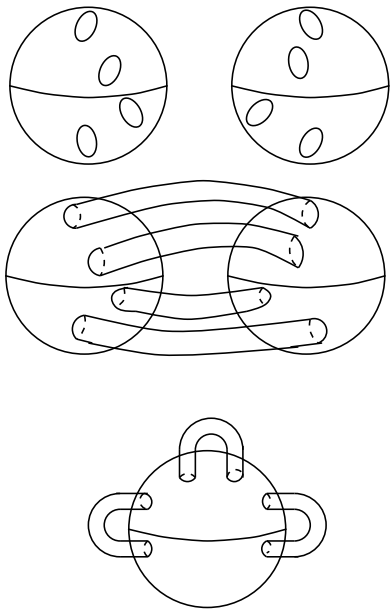

One can find a relation between the degree $h$ of the defining polynomial $P_h(h)$ and the genus $g$ of the surface. In particular

$$g = \begin{cases} \frac{h-2}{2}, & \text{if } h = \text{even}, \\ \frac{h-1}{2}, & \text{if } h = \text{odd}. \end{cases}$$

It is noteworthy that any Riemann surface of genus $g = 1$ or $g = 2$ has a hyper-elliptic representation (2.4). On the other hand, there are surfaces of genus $g \geq 3$ which do not admit a hyper-elliptic representation.

## 3  General Approach to Constructing Integrable Models

By definition, an integrable model of quantum field theory is a model which is characterized by an infinite number of conservation laws.

The most interesting integrable models are the massive ones, i.e. the models that contain in their spectrum massive particles. If we consider the ultraviolet limit of these theories, i.e. the limit of large momenta, the masses of the particles can be ignored and the theories become effectively massless. Massless theories, on the other hand, are scale invariant and in two dimensions, this scale invariance is automatically (under very mild assumptions) extended to an infinite dimensional symmetry. Therefore, any two dimensional massless theory is an integrable theory. Moreover, a classification of all massless field theories and their operator content is possible [2].

The above observation can be used to build integrable massive models in the following way. Given a massless theory with action $S_0$ and some operator content $\{\Phi_i\}$, we define a new model with action

$$S = S_0 + \lambda \int d^2x\, \Phi \,, \tag{3.7}$$

where $\Phi$ is an operator in $\{\Phi_i\}$. The perturbation $\int d^2x\,\Phi$, in general, destroys the conservation laws of the massless theory. However, sometimes it is possible to choose the operaor $\Phi$ in such a way that an infinite subset of the original conservation laws survive in the new theory. There is a well known method [7], known as the *counting argument*, that establishes which conservation laws survive in the new theory.

# 4 Integrable Models

## 4.1 On the Sphere

On the sphere, there is an infinite series of unitary models characterized by an integer $p$ with central charge $c < 1$. The main (primary) operators of these models $\Phi_{mn}$ are classified by the two integers $1 \leq n \leq p$, $1 \leq m \leq p - 1$; all rest operators can be derived from these basic operators. If $S_0(p)$ is the action of the $p$-th unitary model, one can consider, the perturbed models

$$S(p) = S_0(p) + \int d^2x\, \Phi_{nm} \,. \tag{4.8}$$

Obviously, not all of these perturbations will lead to massive integrable models of quantum field theory – if any at all. Using the counting argument for the interesting models, one can show [7] that the following series of models are integrable:

$$S(p) = S_0(p) + \int d^2x\, \Phi_{12} \,, \tag{4.9}$$

$$S(p) = S_0(p) + \int d^2x\, \Phi_{21} \,, \tag{4.10}$$

$$S(p) = S_0(p) + \int d^2x\, \Phi_{13} \,. \tag{4.11}$$

## 4.2 On the Hyper-Elliptic Surfaces

On the hyper-elliptic surfaces, there is also an infinite series of unitary models [4] characterized by an integer $p$ with central charge $c < 2$. The main (primary) operators of these models belong to two categories: the fields $\Phi_{mn}^{m'n'}$ which carry twice as many indices relative to the sphere due to the doubling of the sheets and the fields $\Xi_{nm}$ which carry all the information about the branch cuts of the huper-elliptic surfaces. If $S_0(p)$ is the action

of the $p$-th unitary model on the hyper-elliptic surface, one can consider, the perturbed models

$$S(p) = S_0(p) + \int d^2x \, \Xi_{nm} \, . \tag{4.12}$$

As before, not all of these perturbations will lead to interesting massive integrable models of quantum field theory. Using the counting argument, one can show [1] that the following series of models is integrable:

$$S(p) = S_0(p) + \int d^2x \, \Xi_{12} \, . \tag{4.13}$$

# 5  Other Issues

During the last years, it has been realized that integrable models are characterized by specific symmetries, known as *affine quantum group symmetries*. These symmetries are genereted by non-local conserved currents which in many cases can be constructed explicitly. Such an approach to the quantum field theory permits to obtain non-perturbative solutions, and in particular the $S$-matrix, in the quantum field theory using algebraic methods. This construction will not be explained here, but we refer the interested reader to [1] to see what is known for the present case. In the same reference, he will find a brief discussion about the beta function of the model which confirms the massive spectrum of the model.

# A  Calculational Details

For completeness, in this section we present the mathematical implementation for the construction of the integrable models we described above.

## A.1  CFT on Hyper-Elliptic Surfaces

Let $A_a$, $B_a$, $a = 1, 2, \ldots, g$ be the basic cycles of the hyper-elliptic surface defined by (2.4). As we encircle the point $r_i$ along the contours $A_a$, $B_a$, in the case of an $A_a$ cycle we stay on the same sheet, while in the case of a $B_a$ cycle we pass from the $l$-th sheet to the $(l+1)$-th one. We shall denote the process of encircling the points $w_i$ on the cycles $A_a$, $B_a$ by the symbols $\hat{\pi}_{A_a}$, $\hat{\pi}_{B_a}$ respectively. Here these generators form a group of monodromy that in our case of two-sheet covering of the sphere coincides with the $\mathbb{Z}_2$ group.

We consider the energy-momentum tensor with representation $T^{(l)}(z)$ on each of these sheets. The above definition of the monodromy properties along the cycles $A_a$, $B_a$ implies that the following boundary conditions should be satisfied by the energy-momentum tensor:

$$\hat{\pi}_{A_a} T^{(l)} = T^{(l)}, \quad \hat{\pi}_{B_a} T^{(l)} = T^{(l+1)} \, . \tag{A.14}$$

It is convenient to pass to a basis, in which the operators $\hat{\pi}_{A_a}$, $\hat{\pi}_{B_a}$ are diagonal

$$T = T^{(0)} + T^{(1)} \, , \qquad T^- = T^{(0)} - T^{(1)} \, , \tag{A.15}$$

$$\hat{\pi}_{A_a} T = T \, , \qquad \hat{\pi}_{A_a} T^- = T^- \, , \tag{A.16}$$

$$\hat{\pi}_{B_a} T = T \, , \qquad \hat{\pi}_{B_a} T^- = -T^- \, . \tag{A.17}$$

The corresponding operator product expansions (OPEs) of the $T$, $T^-$ fields can be determined by taking into account the OPEs of $T^{(l)}$, $T^{(l')}$. On the same sheet, the OPEs of the two fields $T^{(l)}(z_1) T^{(l)}(z_2)$, are the same as that on the sphere, while on different sheets they do not correlate, i.e. $T^{(l)}(z_1)T^{(l+1)}(z_2) \sim$ reg. Thus, in the diagonal basis the OPEs can be found to be

$$T(z_1)T(z_2) = \frac{c}{2\,z_{12}^4} + \frac{2\,T(z_2)}{z_{12}^2} + \frac{T'(z_2)}{z_{12}} + \text{reg}\ , \qquad (A.18)$$

$$T^-(z_1)T^-(z_2) = \frac{c}{2\,z_{12}^4} + \frac{2\,T(z_2)}{z_{12}^2} + \frac{T'(z_2)}{z_{12}} + \text{reg}\ , \qquad (A.19)$$

$$T(z_1)T^-(z_2) = \frac{2}{z_{12}^2}T^-(z_2) + \frac{T'^-(z_2)}{z_{12}} + \text{reg}\ , \qquad (A.20)$$

where $c = 2\hat{c}$, and $\hat{c}$ is the central charge in the OPE of $T^{(l)}(z_1)T^{(l)}(z_2)$. It is seen from (A.20) that $T^-$ is primary field with respect to $T$. To write the algebra (A.18)-(A.19) in the graded form we determine the mode expansion of $T$ and $T^-$:

$$T(z)V_{(k)}(0) = \sum_{n \in \mathbb{Z}} z^{n-2} L_{-n} V_{(k)}(0)\ , \qquad (A.21)$$

$$T^-(z)V_{(k)}(0) = \sum_{n \in \mathbb{Z}} z^{n-2-k/2} L^-_{-n+k/2} V_{(k)}(0)\ , \qquad (A.22)$$

where $k$ ranges over the values 0,1 and determines the parity sector in conformity with the boundary conditions (A.16) and (A.17). Standard calculations lead to the following algebra for the operators $L_{-n}$ and $L^-_{-n+k/2}$:

$$\begin{aligned}
[L_n, L_m] &= (n-m)L_{n+m} + \frac{c}{12}(n^3 - n)\delta_{m+n,0}\ , \\
[L^-_{m+k/2}, L^-_{n+k/2}] &= (m-n)L_{n+m+k} + \frac{c}{12}[(m+k/2)^3 - (m+k/2)]\delta_{n+m+k,0}\ , \\
[L_m, L^-_{n+k/2}] &= [m-n-k/2]L^-_{m+n+k/2}\ .
\end{aligned} \qquad (A.23)$$

The operators $\overline{L}_n$, $\overline{L}_{m+k/2}$, $\overline{L}^-_n$, $\overline{L}^-_{m+k/2}$ satisfy the same relations and $\overline{L}_n, \overline{L}_{m+k/2}, \overline{L}^-_n, \overline{L}^-_{m+k/2}$ commute with $L_n$, $L_{m+k/2}$, $L^-_n$, $L^-_{m+k/2}$.

To describe the representations of the algebra (A.23), it is necessary to consider separately the non-twisted sector with $k = 0$ and the twisted sector sector with $k = 1$. In order to write the $[V_{(k)}]$ representation of the algebra (A.23) in a more explicit form, it is convenient to consider the highest weight states. In the $k = 0$ sector, the highest weight state $|\Delta, \Delta^-\rangle$ is determined with the help of a primary field $V_{(0)}$ by means of the formula

$$|\Delta, \Delta^-\rangle = V_{(0)}|\emptyset\rangle\ . \qquad (A.24)$$

Using the definition of vacuum, it is easy to see that

$$\begin{aligned}
L_0|\Delta, \Delta^-\rangle &= \Delta|\Delta, \Delta^-\rangle\ , \quad L^-_0|\Delta, \Delta^-\rangle = \Delta^-|\Delta, \Delta^-\rangle\ , \\
L_n|\Delta, \Delta^-\rangle &= L^-_n|\Delta, \Delta^-\rangle = 0, \quad n \geq 1\ .
\end{aligned} \qquad \begin{aligned}(A.25) \\ (A.26)\end{aligned}$$

In the $k = 1$ sector, we define the vector of highest weight $|\Delta\rangle$ of the algebra to be

$$|\Delta\rangle = V_{(1)}|\emptyset\rangle\ , \qquad (A.27)$$

where $V_{(1)}$ is a primary field with respect to $T$. In analogy with the non-twisted sector we obtain

$$L_0|\Delta\rangle = \Delta|\Delta\rangle, \quad L_n|\Delta\rangle = L^-_{n-1/2}|\Delta\rangle = 0, \quad n \geq 1\ . \qquad (A.28)$$

Thus, the Verma module over the algebra (A.23) is obtained by the action of any number of $L_{-m}$ and $L^-_{-m+k/2}$ operators with $n, m > 0$ on the states (A.24) and (A.27). As was shown in ref. [4] by means of

GKO (coset construction) method, the central charge of a reducible unitary representation of the algebra (A.23) has the form

$$c = 2 - \frac{12}{p(p+1)} = 2\hat{c} \ , \quad p = 3, 4, \ldots \ . \tag{A.29}$$

Using ref. [5], Dotsenko and Fateev [6] gave the complete solution for the minimal model correlation functions on the sphere. They were able to write down the integral representation for the conformal blocks of the chiral vertices in terms of the correlation functions of the vertex operators of a free bosonic scalar field $\Phi$ coupled to a background charge $\alpha_0$. This construction has become known as the Coulomb Gas Formalism (CGF). In the present case, this approach is also applicable by considering a Coulomb gas for each sheet separately but coupled to the same bouckground charge:

$$T^{(l)} = -\frac{1}{2}(\partial_z \Phi^{(l)})^2 + i\alpha_0 \partial_z^2 \Phi^{(l)} \ , \quad \langle \Phi^{(l)}(z)\Phi^{(l')}(z')\rangle = -\delta^{ll'} \ln|z-z'|^2 \ , \tag{A.30}$$

$$\hat{\pi}_{A_a}\partial_z\Phi^{(l)} = \partial_z\Phi^{(l)} \ , \quad \hat{\pi}_{B_a}\partial_z\Phi^{(l)} = \partial_z\Phi^{(l+1)} \ , \tag{A.31}$$

where $c = 2 - 24\alpha_0^2$ or $\alpha_0^2 = 1/2p(p+1)$.

Passing to the basis which diagonalizes the operators $\hat{\pi}_{A_a}$, $\hat{\pi}_{B_a}$, i.e.

$$\begin{aligned}\Phi &= \Phi^{(0)} + \Phi^{(1)} \ , \quad \Phi^- = \Phi^{(0)} - \Phi^{(1)} \ , \\ \hat{\pi}_{A_a}\partial_z\Phi &= \partial_z\Phi \ , \quad \hat{\pi}_{B_a}\partial_z\Phi = \partial_z\Phi \ , \\ \hat{\pi}_{A_a}\partial_z\Phi^- &= \partial_z\Phi^- \ , \quad \hat{\pi}_{B_a}\partial_z\Phi^- = -\partial_z\Phi^- \ ,\end{aligned} \tag{A.32}$$

we finally obtain the bosonization rule for the operators $T$, $T^-$ in the diagonal basis

$$T = -\frac{1}{4}(\partial_z\Phi)^2 + i\alpha_0\partial_z^2\Phi - \frac{1}{4}(\partial_z\Phi^-)^2 \ , \tag{A.33}$$

$$T^- = -\frac{1}{2}\partial_z\Phi\partial_z\Phi^- + i\alpha_0\partial_z^2\Phi^- \ . \tag{A.34}$$

In conventions of ref. [4], the vertex operator with charges $\alpha, \beta$ in the $k = 0$ (non-twisted) sector is given by

$$V_{\alpha\beta}(z) = e^{i\alpha\Phi + i\beta\Phi^-} \ , \tag{A.35}$$

with conformal weights $\Delta = \alpha^2 - 2\alpha_0\alpha + \beta^2$ and $\Delta^- = 2\alpha\beta - 2\alpha_0\beta$.

In the $k = 1$ (twisted) sector the situation is slightly different. Here we have an antiperiodic bosonic field $\Phi^-$, i.e. $\Phi^-(e^{2\pi i}z) = -\Phi^-$; this leads to the deformation of the geometry of space-time. If we recall that the circle is parametrized by $\Phi^- \in S^1[0, 2\pi R]$, the condition $\Phi^- \sim -\Phi^-$ means that pairs of points of $S^1$ have been identified. Thus, $\Phi^-$ lives on the orbifold $S^1/\mathbb{Z}_2$; under the identification $\Phi^- \sim -\Phi^-$ the two points $\Phi^- = 0$ and $\Phi^- = \frac{1}{2}(2\pi R)$ are fixed points. One can try to define the twist fields $\sigma_\epsilon(z)$, $\epsilon = 0, 1$, for the bosonic field $\Phi^-$, with respect to which $\Phi^-$ is antiperiodic. Notice that there is a separate twist field for each fixed point. The OPE of the current $I^- = i\partial_z\Phi^-$ with the field $\sigma_\epsilon$ is then

$$I^-(z)\sigma_\epsilon(0) = \frac{1}{2}z^{-1/2}\hat{\sigma}_\epsilon(0) + \ldots \ . \tag{A.36}$$

The twist fields $\sigma_\epsilon$ and $\hat{\sigma}_\epsilon$ are primary fields for the $T_{\text{orb}} = -\frac{1}{4}(\partial_z\Phi^-)^2$ with dimensions $\Delta_\epsilon = 1/16$ and $\hat{\Delta}_\epsilon = 9/16$ respectively. So, in the twisted sector the highest weight vectors (or primary fields) can be written as follows

$$V_{\gamma\epsilon}^{(t)} = e^{i\gamma\Phi}\sigma_\epsilon \ , \quad \Delta^{(t)} = \gamma^2 - 2\alpha_0\gamma + \frac{1}{16} \ . \tag{A.37}$$

In ref. [4], the anomalous dimensions of the primary fields of the minimal models for the algebra (A.23) were obtained both in the non-twisted and twisted sectors in conformity with the spectrum of the central charge (A.29); in particular, it was found that the charges $\alpha, \beta, \gamma$ of the primary fields corresponding to $k = 0$ and $k = 1$ sectors have the form:

$$\begin{aligned}\alpha_{n'm'}^{nm} &= \tfrac{2-n-n'}{2}\alpha_+ + \tfrac{2-m-m'}{2}\alpha_- \ , \quad \beta_{n'm'}^{nm} = \tfrac{n-n'}{2}\alpha_+ + \tfrac{m-m'}{2}\alpha_- \ , \\ \gamma_{nm} &= \tfrac{2-n}{2}\alpha_+ + \tfrac{2-m}{2}\alpha_- \ ,\end{aligned} \tag{A.38}$$

where $1 \leq n, n' \leq p$, $1 \leq m, m' \leq p-1$, and the constants $\alpha_{\pm}$ are expressed in terms of the background charge $\alpha_0$:

$$\alpha_{\pm} = \alpha_0/2 \pm \sqrt{\alpha_0^2/4 + 1/2} \ . \tag{A.39}$$

We denote the corresponding fields by $V_{n'm'}^{nm}$, $V_{nm}^{(t)}$ and their conformal weights by $\Delta_{n'm'}^{nm}$, $\Delta_{nm}^{(t)}$.

We can thus represent the CFT on a hyper-elliptic surface as a CFT on the plane with an additional symmetry, exactly as described by the algebra (A.23). The corresponding highest weight vectors of the algebra are given by (A.35) and (A.37); finally, the central charge is given by (A.29).

We will confine ourselves to the minimal models on hyper-elliptic surfaces as presented above; keeping this in mind we pass to the construction of perturbed models of these CFTs.

## A.2  Perturbation by $V_{nm}^{(t)}$ and Integrals of Motion

Let $S_p$ be the action the $p$-th conformal minimal model on the hyper-elliptic surface $\Gamma$

$$S_p[\Phi, \Phi^-] \sim \int d^2z \, (\partial_z \Phi \partial_{\bar{z}} \Phi - i\alpha_0 R\Phi) + \int d^2z \, \partial_z \Phi^- \partial_{\bar{z}} \Phi^- \ . \tag{A.1}$$

We now consider the perturbation of this conformal field theory by the degenerate relevant operator $V_{nm}^{(t)}$.

$$S_\lambda = S_p[\Phi, \Phi^-] + \lambda \int d^2z \, e^{i\gamma_{nm} \Phi(z,\bar{z})} \sigma_\epsilon(z, \bar{z}) \ . \tag{A.2}$$

The parameter $\lambda$ is a coupling constant with conformal weight $(1 - \Delta_{nm}^{(t)}, 1 - \Delta_{nm}^{(t)})$.

Obviously, for a generic perturbation the new action $S_\lambda$ does not describe an integrable model. We are going to choose the perturbation in a way that the corresponding field theory is integrable. To prove the integrability of this massive (this claim is proved at the end of the present section) theory, one must calculate the characters of the modules of the identity $I$ and $V_{nm}^{(t)}$.

The "basic" currents $T(z)$ and $T^-(z)$ generate an infinite-dimensional vector subspace $\Lambda$ in the representation space. This subspace can be constructed by successive applications of the generators $L_{-n}$ and $L_{-m}^-$ with $n, m > 0$ to the identity operator $I$. $\Lambda$ can be decomposed to a direct sum of eigenspaces of $L_0$, i.e.

$$\Lambda = \bigoplus_{s=0}^{\infty} \Lambda_s \ , \quad L_0 \, \Lambda_s = s \, \Lambda_s \ . \tag{A.3}$$

The space $\Lambda$ contains the subspace $\Lambda' = \partial_z \Lambda$. Therefore, in order to separate the maximal linearly independent set, one must take the factor space $\hat{\Lambda} = \Lambda/(L_{-1}\Lambda \bigoplus L_{-1}^-\Lambda)$ instead of $\Lambda$. The space $\hat{\Lambda}$ admits a similar decomposition as a direct sum of eigenspaces of $L_0$. It follows that the formula of the character for $\hat{\Lambda}$ takes the form

$$\chi_0 = (1-q)^2 \prod_{n=1}^{+\infty} \frac{1}{(1-q^n)^2} \ . \tag{A.4}$$

The dimensionalities of the subspaces $\hat{\Lambda}_s$ can be determined from the character formula

$$\sum_{s=0}^{\infty} q^s \, \dim(\hat{\Lambda}_s) = (1-q) \chi_0 + q \ . \tag{A.5}$$

On the other hand, the module $V$ of the primary field $V_{nm}^{(t)}$ can be constructed by successively applying the generators $L_{-k}$ and $L_{1/2-l}^-$ with $k, l > 0$ to the primary field $V_{nm}^{(t)}$. This space $V$ and the corresponding factor space $\hat{V} = V/L_{-1}V$ may also be decomposed in a direct sum of $L_0$ eigenspaces:

$$V = \bigoplus_{s=0}^{\infty} V_s^{(t)} \ , \quad L_0 \, V_s^{(t)} = s \, V_s^{(t)} \ . \tag{A.6}$$

The dimensionalities of $\widehat{V}_s^{(t)}$ in this factor space associated with the relevant field $V_{(1,1)}^{(t)} = e^{i\frac{\alpha_0}{2}\Phi}\sigma_\epsilon$ are given by the character formula

$$\sum_{s=\mathbb{N}/2}^{+\infty} q^{s+\Delta_{(1,1)}^{(t)}} \dim(\widehat{V}_s^{(t)}) = \chi_{\Delta_{(1,1)}^{(t)}}(1-q) \,, \qquad (A.7)$$

where

$$\chi_{\Delta_{(1,1)}^{(t)}} = q^{\Delta_{(1,1)}^{(t)}} \prod_{n=1}^{+\infty} \frac{1}{(1-q^n)(1-q^{n-1/2})} \,, \quad \Delta_{(1,1)}^{(t)} = \frac{1}{16}\left(1 - \frac{6}{p(p+1)}\right) \,. \qquad (A.8)$$

When the dimensionalities of $\widehat{V}_s^{(t)}$ (calculated from (A.7), (A.8)) are compared to those of $\widehat{\Lambda}_{s+1}$, we see that for $s = 1, 3, 5, \ldots$ the $\dim(\widehat{\Lambda}_{s+1})$ exceeds the $\dim(\widehat{V}_s^{(t)})$ at least by the unity, i.e. $\dim(\widehat{\Lambda}_{s+1}) > \dim(\widehat{V}_s^{(t)})$, $s = 1, 3, 5, \ldots$. This proves that the model

$$S_\lambda = S_p + \lambda \int d^2 z \, e^{i\frac{\alpha_0}{2}\Phi(z,\overline{z})} \sigma_{\epsilon'}(z,\overline{z}) \qquad (A.9)$$

possesses an infinite set of non-trivial IMs. We note here that there are no such IMs for perturbations by the operators $V_{nm}^{(t)}$ with $n, m > 1$.